\documentclass[12pt]{article}
\usepackage{axodraw}
\usepackage{psfig}
\newcommand{\be}{\begin{equation}}
\newcommand{\ee}{\end{equation}}
\newcommand{\sss}[1]{\mbox{\scriptsize #1}}

\newcommand{\real}{{\cal\mbox{Re\,}}}

\newcommand{\M}{{\cal M}}
\newcommand{\J}{{\cal J}}
\newcommand{\I}{{\cal I}}

\newcommand{\OO}{{\cal O}}

\newcommand{\lsim}
{\;\raisebox{-.3em}{$\stackrel{\displaystyle <}{\sim}$}\;}
\newcommand{\gsim}
{\;\raisebox{-.3em}{$\stackrel{\displaystyle >}{\sim}$}\;}

\newcommand{\GeV}{\unskip\,\mathrm{GeV}}
\newcommand{\MeV}{\unskip\,\mathrm{MeV}}

\begin{document}
\pagestyle{empty}
\begin{flushright}
{DTP/99/14}\\
{INLO-PUB 2/99}
\end{flushright}
\vspace*{5mm}
\begin{center}
  {\bf 
   ONE-LOOP QCD INTERCONNECTION EFFECTS IN PAIR PRODUCTION OF TOP QUARKS
  } \\
\vspace*{1cm} 
{\bf W.~Beenakker}$^{*)}$\\ 
\vspace{0.3cm}
Physics Department, University of Durham, Durham DH1 3LE, England\\
\vspace{0.5cm}
{\bf F.A.~Berends} \ \  
{\bf and}  \ \ 
{\bf A.P.~Chapovsky}$^{\dagger)}$\\
\vspace{0.3cm}
Instituut--Lorentz, University of Leiden, P.O.~Box 9506,  2300 RA Leiden,  The Netherlands\\
\vspace*{2cm}  
                                {\bf ABSTRACT} \\ \end{center}
\vspace*{5mm}
\noindent
        We calculate the one-loop non-factorizable QCD corrections to the
        production and decay of pairs of top quarks at various collider
        experiments. These non-factorizable corrections interconnect the
        different production and decay stages of the off-shell top-pair 
        production processes. This in particular affects the invariant-mass 
        distributions of the off-shell top quarks, resulting in a
        shift of the maximum of the distorted Breit--Wigner distributions.
        Although the non-factorizable corrections can be large, the 
        actual shift in the mass as determined from the peak position
        of the corrected Breit--Wigner line-shape is below $100\MeV$.
	\\
\vspace*{3cm}\\ 
\begin{flushleft}
February 1999
\end{flushleft}
\noindent 
\rule[.1in]{16.5cm}{.002in}

\noindent
$^{*)}$Research supported by a PPARC Research Fellowship.\\
$^{\dagger)}$Research supported by the Stichting FOM.
\vspace*{0.3cm}

\vfill\eject

\setcounter{page}{1}
\pagestyle{plain}

\section{Introduction}

At present and future collider experiments, a detailed investigation of the 
production of top-quark pairs will substantially contribute to our knowledge 
of the top-quark properties and thereby of the  Standard Model. 
An improved measurement of the top-quark mass 
$m_t$, for instance, can serve to obtain improved indirect sensitivity to the 
mass of the Standard Model Higgs boson. This is achieved by combining the 
high-precision measurements of the electroweak parameters at LEP/SLC with the 
direct measurements of the top-quark and $W$-boson masses. 

Pairs of top quarks can be produced in hadron collisions at the Tevatron  
($p\bar{p}$) and LHC ($pp$), as well as in $e^+e^-$ and $\gamma\gamma$
collisions at a future linear collider. Since the top quark has a large width 
as compared to the QCD hadronization scale, 
$\Gamma_{t} \approx 1.4 \GeV \gg \Lambda_{\sss{QCD}} \approx 200-300\MeV$, it 
predominantly decays before hadronization takes place. Therefore the 
perturbative approach can be used for describing top quarks. 
The main lowest-order (partonic) mechanisms for the pair production of
top quarks are
\be
\label{ee->tt}
  e^+e^-,\gamma\gamma \to t\bar{t} \to bW^+\bar{b}W^- \to 6\ \mbox{fermions},
\ee 
\be
\label{qq->tt}
  q\bar{q},gg \to t\bar{t} \to bW^+\bar{b}W^- \to 6\ \mbox{fermions}.
\ee
A lot of effort has been put into an adequate theoretical description of these
reactions (see e.g.~Ref.~\cite{tt-review} for two review papers). Most of 
these studies 
treat the top quarks as stable particles, which is a reasonable approximation 
since $\Gamma_t/m_t = \OO(1\%)$. For the reactions $q\bar{q},gg \to t\bar{t}$ 
these studies comprise QCD~\cite{tt-hadprod-QCD} and 
electroweak~\cite{tt-hadprod-ew} one-loop corrections, as well as the 
resummation of soft-gluon effects~\cite{softgluon}. Also for the reactions 
$e^+e^-,\gamma\gamma \to t\bar{t}$ both the QCD~\cite{tt-LCprod-QCD} and 
electroweak~\cite{tt-LCprod-ew} one-loop corrections are known. Moreover,
the $t\bar{t}$ threshold, with its sizeable QCD~\cite{tt-threshold} 
and Yukawa interactions~\cite{yukawa}, 
has been analysed in detail.

One would, however, like to treat the top quark as an unstable particle, with 
a Breit--Wigner distribution describing its line shape. The most economic
approach for treating processes that involve the production and subsequent  
decay of unstable particles is the so-called leading-pole approximation 
(LPA)~\cite{pole-scheme}. This approximation is based on an expansion of the 
complete 
amplitude around the poles of the unstable particles, which can be viewed as a 
prescription for performing an effective expansion in powers of $\Gamma_i/M_i$.
Here $M_i$ and $\Gamma_i$ stand for the masses and widths of the various 
unstable particles. The residues in the pole expansion are physically 
observable and therefore gauge-invariant. The actual approximation consists in
retaining only the terms with the highest degree of resonance. In the case of 
top-quark pair production only the double-pole residues are hence considered
and the LPA becomes a double-pole approximation (DPA). This approximation will 
be valid sufficiently far above the $t\bar{t}$-threshold.
If in reactions (\ref{ee->tt}) and (\ref{qq->tt}) also the $W$ bosons are
treated as unstable particles, then also for these particles the leading
pole residues should be taken. In this approach the complete set of 
corrections to reactions (\ref{ee->tt}) and (\ref{qq->tt}) naturally
splits into two groups: factorizable and non-factorizable corrections. The
factorizable corrections are directly linked to the density matrices for
on-shell production and decay of the unstable particles. The non-factorizable 
corrections can be viewed as describing interactions that interconnect
different (production/decay) stages of the off-shell process. A detailed 
discussion of this method with all its subtleties can be found in 
Ref.~\cite{ww-dpa}, where the method has been applied to the complete set of 
$\OO(\alpha)$ radiative corrections to the process 
$e^+e^- \to W^+W^- \to 4\ \mbox{fermions}$.
For $t\bar{t}$ production partial results along this line 
exist~\cite{tt-dpa}, involving a subset of the factorizable corrections to the 
reaction $e^+e^- \to t\bar{t} \to bW^+\bar{b}W^- \to 6\ \mbox{fermions}$. 
However, the non-factorizable corrections are needed for a complete 
$\OO(\alpha_{\sss{s}})$ calculation.

In recent years the necessary methods for calculating such non-factorizable 
corrections have been developed. In Ref.~\cite{nf-my} a first complete
calculation was performed for $t\bar{t}$ production in $e^+e^-$ collisions.
In Ref.~\cite{nf-bbc} the non-factorizable corrections were calculated for  
$W$-pair production, revealing differences with the results of 
Ref.~\cite{nf-my}. The results of Ref.~\cite{nf-bbc} were confirmed by  
an independent calculation~\cite{nf-ddr} as well as by a re-analysis of the 
results of Ref.~\cite{nf-my}.

In this paper we apply our calculations presented in Ref.~\cite{nf-bbc}
to the non-factorizable $\OO(\alpha_{\sss{s}})$ corrections to $t\bar{t}$
production at various colliders. 
We discuss the effect on the invariant-mass distribution of the 
off-shell top quark and the resulting shift in the maximum of the distorted 
Breit--Wigner distribution.

\section{Definition of the non-factorizable corrections}

In the LPA approach reactions like (\ref{ee->tt}) and (\ref{qq->tt}), which 
involve unstable particles during intermediate stages, can be viewed as 
consisting of separate subprocesses, i.e.~the production and decay of the 
unstable particles. Having this picture in mind, the complete set of radiative 
corrections can be separated naturally into a sum of corrections to these 
subprocesses, called factorizable corrections, and those corrections that 
interconnect various subprocesses, called non-factorizable corrections. 
It should be noted, however, that it is often misleading to identify the 
non-factorizable contributions on the basis of diagrams. Such a
definition is in general not gauge-invariant. Rather one should realize that 
only real/virtual semi-soft gluons%
\footnote{These gluons will still be perturbative
          in our case as their typical energy 
          ($E_g \sim \Gamma_{t,W} \gsim 1.4\GeV$) largely exceeds the QCD
          hadronization scale ($\Lambda_{\sss{QCD}} \approx 200-300\MeV$).}
with $E_g = \OO(\Gamma_i)$ will contribute, the contributions of the hard 
gluons being suppressed by $\Gamma_i/E_g$. This is a consequence of the fact 
that the various subprocesses are typically separated by a big space-time 
interval of $\OO(1/\Gamma_i)$ due to the propagation of the unstable particles.
The subprocesses can be interconnected only by the radiation of semi-soft 
gluons with energy of $\OO(\Gamma_i)$, which induce interactions that are
sufficiently long range. Hard gluons ($E_g = \OO(M_i) \gg \Gamma_i$) as well
as massive particles induce short-range interactions and therefore contribute 
exclusively to the factorizable corrections, which are governed by the
relatively short time interval $\sim 1/M_i$ on which the decay and production
subprocesses occur. A more detailed discussion of these issues can be found 
in Refs.~\cite{ww-dpa,dkos,nf-cancellations}.

\begin{figure}
  \begin{center}
  \begin{picture}(200,120)(0,0)
    \ArrowLine(43,58)(25,40)        \Text(7,40)[lc]{$\bar{q}$}
    \ArrowLine(25,90)(43,72)        \Text(7,92)[lc]{$q$}
    \ArrowLine(50,65)(100,90)       \Text(75,90)[]{$t$}   
    \ArrowLine(100,90)(150,115)     \Text(125,115)[]{$b$}  
    \ArrowLine(100,40)(50,65)       \Text(75,40)[]{$\bar{t}$}  
    \ArrowLine(150,15)(100,40)      \Text(125,15)[]{$\bar{b}$}
    \Photon(100,90)(150,90){1}{5}   \Text(125,80)[]{$W^{+}$}
    \Photon(100,40)(150,40){1}{5}   \Text(125,50)[]{$W^{-}$}
    \ArrowLine(150,90)(200,110)      
    \ArrowLine(200,70)(150,90)    
    \ArrowLine(150,40)(200,60)       
    \ArrowLine(200,20)(150,40)    
    \Gluon(100,90)(100,40){3}{7}   \Text(92,65)[]{$g$}
    \GCirc(50,65){10}{1}  
    \GCirc(100,90){5}{1}
    \GCirc(100,40){5}{1} 
    \GCirc(150,90){5}{1}
    \GCirc(150,40){5}{1}
  \end{picture}
  \end{center}
  \caption[]{The generic structure of the complete $t\bar{t}$-production 
             process 
             $q\bar{q} \to t\bar{t} \to bW^+\bar{b}W^- \to 6\ \mbox{fermions}$ 
	     in the LPA. The open circles denote the 
             various production and decay subprocesses. As an example also the
             non-factorizable semi-soft gluon interaction between the two 
             top-quark decay subprocesses is shown.}       
  \label{fig:diag}
\end{figure}
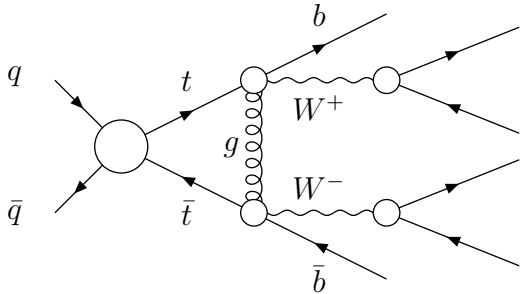

In Fig.~\ref{fig:diag} we show schematically the partonic process 
$q\bar{q} \to t\bar{t} \to bW^+\bar{b}W^- \to 6\ \mbox{fermions}$. The process
consists of five subprocesses, which we will denote by $t\bar{t}_{\sss{prod}}$,
$t_{\sss{dec}}$, $\bar{t}_{\sss{dec}}$, $W^{+}_{\sss{dec}}$, and 
$W^{-}_{\sss{dec}}$. In Fig.~\ref{fig:diag} these subprocesses are indicated 
by the open circles. The non-factorizable semi-soft gluon interactions 
interconnect any two different subprocesses, as is exemplified in 
Fig.~\ref{fig:diag} for the two top-quark decay subprocesses. The coupling of 
such a gluon to a certain subprocess can be written in terms of semi-soft
currents. In contrast to soft-gluon currents, the effect
of the gluon momentum on the unstable-particle propagators 
cannot be neglected in the semi-soft currents. 
The various non-factorizable corrections to the 
cross-section are just given by all possible interferences of the
semi-soft currents. This will be made more explicit in the next section.

\section{Colour dependence of the non-factorizable corrections}

We start off by considering the simpler case of stable $W$ bosons. At
the end of this section we will indicate what happens if the $W$
bosons decay hadronically. For stable $W$ bosons one can identify three 
subprocesses: $t\bar{t}_{\sss{prod}}$, $t_{\sss{dec}}$, $\bar{t}_{\sss{dec}}$.
The non-factorizable corrections are given by the semi-soft gluon interferences
between these different subprocesses. As only semi-soft gluons contribute, the
virtual and real matrix elements factorize in terms of lowest-order matrix 
elements and semi-soft currents. In view of the possible presence of coloured 
particles in the initial state ($q\bar{q},gg$), this factorization depends on 
the colour structure. For the reactions (\ref{ee->tt}), which involve only 
colourless initial-state particles, the $t\bar{t}$ pair is produced in a 
singlet state. In contrast, the $t\bar{t}$ pair is produced in an octet state 
in the lowest-order annihilation process $q\bar{q} \to t\bar{t}$, which 
involves the time-like exchange of a gluon. Both singlet and octet states are 
present in the lowest-order gluon-fusion reaction $gg \to t\bar{t}$, since in 
that case also space-like top-quark-exchange diagrams contribute. Because of 
these differences in the colour structure of the lowest-order reactions, also 
the non-factorizable corrections will come out differently, as we will see
from the following discussion.

In order to keep the notation as general as possible, we write the 
lowest-order partonic reactions in the generic form  
\be
  Q_1(q_1)Q_2(q_2) \to t(p_1)\bar{t}(p_2) 
                   \to b(k_1)W^+(k_1')\,\bar{b}(k_2)W^-(k_2'),
\ee
where $\,Q_1Q_2 = \{e^+e^-,\gamma\gamma,q\bar{q},gg\}$. The corresponding
lowest-order matrix element will be denoted by $\,(\M_0)_{ij}^{c_2c_1}$,
where $i,j$ indicate the $t,\bar{t}$ colour indices in the fundamental
representation. The colour indices $c_1,c_2$ belonging to $Q_1,Q_2$ depend on
the specific initial state: they are absent for the colourless $e^+e^-$ and 
$\gamma\gamma$ initial states, and they are in the fundamental/adjoint 
representation for the $q\bar{q}/gg$ initial states. The momentum, Lorentz 
index, and colour index of the semi-soft gluon will be denoted by $k$, $\mu$, 
and $a$, respectively.

By using the relation
\be
  (T^a)_{ij}(T^a)_{kl} = \frac{1}{2}\,\biggl( \delta_{il}\,\delta_{kj}
                            - \frac{1}{N}\,\delta_{ij}\,\delta_{kl} \biggr)
\ee
for the $SU(N)$ generators $T^a$ in the fundamental representation (with $N=3$
for QCD), the virtual and real non-factorizable corrections take the generic 
form:
\begin{eqnarray}
\label{nf/virt}
  d\sigma_{\sss{nf}}^{\sss{virt}}
  &=&
  \frac{1}{K_{\sss{in}}}\,\frac{d\Gamma_0}{2s}\,
  (\M_0^*)_{i''j''}^{c_2''c_1''}\,(\M_0)_{i'j'}^{c_2'c_1'}\,
  \real \Biggl\{ i \int\frac{d^4 k}{(2\pi)^4 [k^2+io]}
                 \Bigl(\Delta_{\sss{nf}}^{\sss{virt}}
                 \Bigr)_{i''i';j'j''}^{c_2''c_2';c_1'c_1''}
        \Biggr\}, \\[1mm]
\label{nf/real}
  d\sigma_{\sss{nf}}^{\sss{real}}
  &=&
  {}- \frac{1}{K_{\sss{in}}}\,\frac{d\Gamma_0}{2s}\,
  (\M_0^*)_{i''j''}^{c_2''c_1''}\,(\M_0)_{i'j'}^{c_2'c_1'}\,
  \real \Biggl\{ \int\frac{d\vec{k}}{(2\pi)^3 2k_0}
                 \Bigl(\Delta_{\sss{nf}}^{\sss{real}}
                 \Bigr)_{i''i';j'j''}^{c_2''c_2';c_1'c_1''}
        \Biggr\}.
\end{eqnarray}
Here the pre-factor consists of the lowest-order phase-space factor in the 
LPA [$d\Gamma_0$], the partonic flux factor [$1/(2s)$], and the initial-state 
spin and colour average [$1/K_{\sss{in}}$]. The non-factorizable kernels can
be expressed in terms of semi-soft currents according to  
\begin{eqnarray}
 \Bigl(\Delta_{\sss{nf}}^{\sss{virt}}
 \Bigr)_{i''i';j'j''}^{c_2''c_2';c_1'c_1''}\!\!\!\!\! 
 &=&\!\!\!
 \frac{1}{2}\,\delta_{c_2''c_2'}\,\delta_{c_1'c_1''}
 \Biggl\{ \delta_{i''j''}\,\delta_{j'i'}
          \biggl[ \J_t^{\mu} (\J_{t\bar{t}} - \tilde{\J}_{t\bar{t}}
                             - \tilde{\J}_{\oplus})_{\mu}
               + \J_{\bar{t}}^{\mu}(\J_{t\bar{t}} + \tilde{\J}_{t\bar{t}}
                             + \tilde{\J}_{\ominus})_{\mu}
               + 2\J_t^{\mu}\J_{\bar{t},\,\mu}
          \biggr] \nonumber \\[1mm]
 & &\!\!\! \hphantom{\frac{1}{2}\delta_{c_2''c_2'}\delta_{c_1'c_1''}\!}
      {}+ \delta_{i''i'}\,\delta_{j'j''}  
          \biggl[ N\J_t^{\mu} (\J_{t\bar{t}} + \tilde{\J}_{t\bar{t}}
                              + \tilde{\J}_{\oplus})_{\mu}
               + N\J_{\bar{t}}^{\mu} (\J_{t\bar{t}}- \tilde{\J}_{t\bar{t}}
                                      - \tilde{\J}_{\ominus})_{\mu}
          \nonumber \\[1mm]
 & &\!\!\! \hphantom{\frac{1}{2}\,\delta_{c_2''c_2'}\,\delta_{c_1'c_1''}
                   \delta_{i''i'}\,\delta_{j'j''}aa} 
             {}- \frac{2}{N}\,(\J_t^{\mu}\J_{\bar{t},\,\mu}
                               + \J_t^{\mu}\J_{t\bar{t},\,\mu}
                               + \J_{\bar{t}}^{\mu}\J_{t\bar{t},\,\mu})
          \biggr] 
 \Biggr\} \nonumber \\[1mm]
 & &\!\!\!  
      {}+ \Bigl(Q_{\sss{in}}^a\Bigr)^{c_2''c_2';c_1'c_1''} 
          \biggl\{ \delta_{j'j''}\,(T^a)_{i''i'}\,\J_t^{\mu}\J_{\oplus,\,\mu}
          + \delta_{i''i'}\,(T^a)_{j'j''}\,\J_{\bar{t}}^{\mu}\J_{\ominus,\,\mu}
          \biggr\}, \\[1mm]
 \Bigl(\Delta_{\sss{nf}}^{\sss{real}}
 \Bigr)_{i''i';j'j''}^{c_2''c_2';c_1'c_1''}\!\!\!\!\! 
 &=&\!\!\!
 \frac{1}{2}\,\delta_{c_2''c_2'}\,\delta_{c_1'c_1''}
 \Biggl\{ \delta_{i''j''}\,\delta_{j'i'}
          \biggl[ \I_t^{*\,\mu} (\I_{t\bar{t}} - \tilde{\I}_{t\bar{t}}
                                - \tilde{\I}_0)_{\mu}
               + \I_{\bar{t}}^{*\,\mu} (\I_{t\bar{t}} + \tilde{\I}_{t\bar{t}}
                                        + \tilde{\I}_0)_{\mu}
               + 2\I_t^{*\,\mu}\I_{\bar{t},\,\mu}
          \biggr] \nonumber \\[1mm]
 & &\!\!\! \hphantom{\frac{1}{2}\delta_{c_2''c_2'}\delta_{c_1'c_1''}\!}
      {}+ \delta_{i''i'}\,\delta_{j'j''}  
          \biggl[ N\I_t^{*\,\mu} (\I_{t\bar{t}} + \tilde{\I}_{t\bar{t}}
                                 + \tilde{\I}_0)_{\mu}
               + N\I_{\bar{t}}^{*\,\mu} (\I_{t\bar{t}} - \tilde{\I}_{t\bar{t}}
                                         - \tilde{\I}_0)_{\mu}
          \nonumber \\[1mm]
 & &\!\!\! \hphantom{\frac{1}{2}\,\delta_{c_2''c_2'}\,\delta_{c_1'c_1''}
                   \delta_{i''i'}\,\delta_{j'j''}aa} 
             {}- \frac{2}{N}\,(\I_t^{*\,\mu}\I_{\bar{t},\,\mu}
                               + \I_t^{*\,\mu}\I_{t\bar{t},\,\mu}
                               + \I_{\bar{t}}^{*\,\mu}\I_{t\bar{t},\,\mu})
          \biggr] 
 \Biggr\} \nonumber \\[1mm]
 & &\!\!\!
      {}+ \Bigl(Q_{\sss{in}}^a\Bigr)^{c_2''c_2';c_1'c_1''} 
          \biggl\{ \delta_{j'j''}\,(T^a)_{i''i'}\,\I_t^{*\,\mu}\I_{0,\,\mu}
          + \delta_{i''i'}\,(T^a)_{j'j''}\,\I_{\bar{t}}^{*\,\mu}\I_{0,\,\mu}
          \biggr\}.
\end{eqnarray}
The terms proportional to $\delta_{i''j''}\delta_{j'i'}$ project on the 
lowest-order singlet $t\bar{t}$ states, whereas the terms proportional to 
$\delta_{i''i'}\,\delta_{j'j''}$ completely factorize the lowest-order
cross-section. The colour structure 
\be
        Q_{\sss{in}}^a  = \left\{
          \begin{tabular}{ll}
            0   
            & 
            for \  $e^{+}e^{-}, \gamma\gamma$
            \\[1mm] 
            $\delta_{c_{1}' c_{1}''} (T^{a})_{c_2'' c_2'} 
            + 
            \delta_{c_{2}'' c_{2}'} (T^{a})_{c_1' c_1''}$ 
            & 
            for \  $q\bar{q}$ 
            \\[1mm]
            $\delta_{c_{1}' c_{1}''} (F^{a})_{c_2'' c_2'} 
            + 
            \delta_{c_{2}'' c_{2}'} (F^{a})_{c_1' c_1''}$ 
            & 
            for \  $gg$
          \end{tabular}
          \right.
\ee
depends on the specific initial state and in general does not project on 
explicit lowest-order $t\bar{t}$ colour states. 
Here $F^{a}$ are the $SU(N)$ generators in the adjoint representation, 
which are defined in terms of the $SU(N)$ structure constant according to
$(F^{a})_{bc}=-if^{abc}$. Note that for $e^{+}e^{-}$ and $\gamma\gamma$
initial states the currents $\tilde{\J}_{\ominus}^{\mu}$, 
$\tilde{\J}_{\oplus}^{\mu}$ and $\tilde{\I}_0^{\mu}$ completely drop out
of Eqs.~(\ref{nf/virt}) and (\ref{nf/real}), as it should be for colourless
particles in the initial state.

The semi-soft currents appearing in the virtual non-factorizable corrections 
are given by 
\be
  \J_t^{\mu}
  = {}- g_s\Biggl[ \frac{p_1^{\mu}}{kp_1+io}
                 - \frac{k_1^{\mu}}{kk_1+io}
           \Biggr]\frac{D_1}{D_1+2kp_1}, 
  \quad\quad
  \J_{\bar{t}}^{\mu}
  = {}- g_s\Biggl[ \frac{p_2^{\mu}}{-kp_2+io}
                 - \frac{k_2^{\mu}}{-kk_2+io}
           \Biggr]\frac{D_2}{D_2-2kp_2}
\ee
for gluon emission from the decay stages of the process, and 
\begin{eqnarray}
  \J_{t\bar{t}}^{\mu} 
  &=&
  g_s\Biggl[ \frac{p_1^{\mu}}{kp_1+io} + \frac{p_2^{\mu}}{-kp_2+io} \Biggr],
  \quad\quad
  \ \ \tilde{\J}_{t\bar{t}}^{\mu} \ =\ g_s\Biggl[ \frac{p_1^{\mu}}{kp_1+io}
                                                - \frac{p_2^{\mu}}{-kp_2+io}
                                          \Biggr], 
  \nonumber \\[1mm]
  \J_{\oplus}^{\mu}
  &=&
  {}- g_s\Biggl[ \frac{q_1^{\mu}}{kq_1+io} - \frac{q_2^{\mu}}{kq_2+io} \Biggr],
  \quad\quad
  \ \tilde{\J}_{\oplus}^{\mu} \ =\ {}- g_s\Biggl[ \frac{q_1^{\mu}}{kq_1+io} 
                                                + \frac{q_2^{\mu}}{kq_2+io} 
                                          \Biggr],
  \nonumber \\[1mm]
  \J_{\ominus}^{\mu}
  &=&
  g_s\Biggl[ \frac{q_1^{\mu}}{-kq_1+io} - \frac{q_2^{\mu}}{-kq_2+io} \Biggr],
  \quad\quad
  \tilde{\J}_{\ominus}^{\mu} \ =\ g_s\Biggl[ \frac{q_1^{\mu}}{-kq_1+io} 
                                           + \frac{q_2^{\mu}}{-kq_2+io} 
                                     \Biggr]
\end{eqnarray}
for gluon emission from the production stage of the process. Here $g_s$ is the 
QCD gauge coupling and $D_{1,2} = p_{1,2}^2 - m_t^2 + im_t\Gamma_t$ is a 
shorthand notation for the inverse top-quark propagators. Note the 
difference in the sign of the $io$ parts appearing in the currents 
$\J_{\oplus},\tilde{\J}_{\oplus}$ and $\J_{\ominus},\tilde{\J}_{\ominus}$. 
These infinitesimal imaginary parts are needed to ensure a proper 
incorporation of causality.
 
The corresponding semi-soft real-gluon currents read
\be
  \I_t^{\mu}
  = - g_s\Biggl[ \frac{p_1^{\mu}}{kp_1}
               - \frac{k_1^{\mu}}{kk_1}
         \Biggr]\frac{D_1}{D_1+2kp_1}, 
  \quad\quad
  \I_{\bar{t}}^{\mu}
  = g_s\Biggl[ \frac{p_2^{\mu}}{kp_2}
             - \frac{k_2^{\mu}}{kk_2}
         \Biggr]\frac{D_2}{D_2+2kp_2}
\ee
and
\begin{eqnarray}
  \I_{t\bar{t}}^{\mu} 
  &=&
  g_s\Biggl[ \frac{p_1^{\mu}}{kp_1} - \frac{p_2^{\mu}}{kp_2} \Biggr],
  \quad\quad
  \ \ \ \tilde{\I}_{t\bar{t}}^{\mu} \ =\ g_s\Biggl[ \frac{p_1^{\mu}}{kp_1}
                                                    + \frac{p_2^{\mu}}{kp_2}
                                              \Biggr], 
  \nonumber \\[1mm]
  \I_0^{\mu}
  &=&
  {}- g_s\Biggl[ \frac{q_1^{\mu}}{kq_1} - \frac{q_2^{\mu}}{kq_2} \Biggr],
  \quad\quad
  \tilde{\I}_0^{\mu} \ =\ {}- g_s\Biggl[ \frac{q_1^{\mu}}{kq_1} 
                                       + \frac{q_2^{\mu}}{kq_2} 
                                 \Biggr].
\end{eqnarray}

By simple power counting one can explicitly see from the above specified 
currents that the contributions of hard gluons are suppressed and that
effectively only semi-soft gluons with $E_g = k_0 = \OO(\Gamma_t)$ contribute.
In view of the pole structure of the virtual corrections, governed by the
infinitesimal imaginary parts $io$, many of the non-factorizable corrections 
will vanish when virtual and real-gluon corrections are added up. For instance,
all initial--final state interferences will vanish, leaving behind a
very limited subset of `final-state' 
interferences~\cite{ww-dpa,nf-cancellations}.
The following holds for the remaining interferences:
$$
  \I_t^{*\,\mu}\tilde{\I}_{t\bar{t},\,\mu} 
  \to 
  - \I_t^{*\,\mu}\I_{t\bar{t},\,\mu},
  \quad\quad
  \I_{\bar{t}}^{*\,\mu}\tilde{\I}_{t\bar{t},\,\mu} 
  \to 
  \I_{\bar{t}}^{*\,\mu}\I_{t\bar{t},\,\mu},
$$
with similar effective replacements for $\tilde{\J}_{t\bar{t}}$. As a result 
of these properties of the non-factorizable corrections, a factorization per 
colour structure emerges:
\begin{eqnarray}
\label{sig_nf}
  d\sigma_{\sss{nf}}
  &=&
  \delta_{\sss{nf}}\,\Biggl[ \frac{N^2-1}{2N}\,d\sigma_{\sss{Born},1}
                           - \frac{1}{2N}\,d\sigma_{\sss{Born},8}
                     \Biggr],
  \\[1mm]
  \delta_{\sss{nf}}
  &=&
  2\real \Biggl\{ i \int\frac{d^4 k}{(2\pi)^4 [k^2+io]}
                    \Bigl[  \J_t^{\mu}\J_{\bar{t},\,\mu} 
                           + \J_t^{\mu}\J_{t\bar{t},\,\mu}
                           + \J_{\bar{t}}^{\mu}\J_{t\bar{t},\,\mu}
                    \Bigr]
  \nonumber \\
  & & \hphantom{2\real a}
                {}- \int\frac{d\vec{k}}{(2\pi)^3 2k_0}
                    \Bigl[  \I_t^{*\,\mu}\I_{\bar{t},\,\mu}
                          + \I_t^{*\,\mu}\I_{t\bar{t},\,\mu}
                          + \I_{\bar{t}}^{*\,\mu}\I_{t\bar{t},\,\mu}
                    \Bigr]
         \Biggr\}.
\end{eqnarray}  
Here $d\sigma_{\sss{Born},1}$ and $d\sigma_{\sss{Born},8}$ are the
lowest-order multi-differential 
cross-sections for producing the intermediate $t\bar{t}$ pair in a
singlet and octet state, respectively. 
For completeness we note that
\be
\label{sig_born}
         d\sigma_{\sss{Born}}^{e^{+}e^{-}, \gamma\gamma}  
         =
         d\sigma_{\sss{Born},1}^{e^{+}e^{-}, \gamma\gamma},
         \ \ \ 
         d\sigma_{\sss{Born}}^{q\bar{q}}  
         =
         d\sigma_{\sss{Born},8}^{q\bar{q}},  
         \ \ \ 
         d\sigma_{\sss{Born}}^{gg}  
         =
         d\sigma_{\sss{Born},1}^{gg}
         +
         d\sigma_{\sss{Born},8}^{gg}.
\ee
The non-factorizable factor $\delta_{\sss{nf}}$ can be obtained from 
Ref.~\cite{nf-bbc}. 
The results of Section~4 of that paper should be used, since
those allow for massive decay products from the unstable particles,
which is the case for the top-quark decay. 

We conclude by considering the case that also the $W$ bosons are unstable.
This adds two decay subprocesses, $W^{+}_{\sss{dec}}$ and $W^{-}_{\sss{dec}}$,
to the three we have considered so far. If the $W$ bosons decay leptonically,
nothing changes as the gluon cannot couple to the $W$ decay subprocesses in
that case. For a hadronically decaying $W$ boson additional interferences have
to be taken into account. However, such interferences trivially vanish as a
result of the singlet nature of the $W$-boson decays [i.e.~Tr($T^a$)=0].

\section{Numerical results}

With the help of  Eqs.~(\ref{sig_nf})--(\ref{sig_born}) we can now 
in principle evaluate all kinds of multi-differential distributions,
with and without non-factorizable corrections.
Although the factorized structure of the non-factorizable corrections is 
very transparent in Eq.~(\ref{sig_nf}), integration of the multi-differential 
cross-sections will affect this structure. For instance, in Eq.~(\ref{sig_nf}) 
the correction to the singlet cross-section differs by a factor $-8$ with 
respect to the octet one. However, for the calculation of the relative
non-factorizable corrections to a one-dimensional distribution, one has to 
evaluate the ratio of the integrated Eqs.~(\ref{sig_nf}) and (\ref{sig_born}). 
Since $\delta_{\sss{nf}}$ depends on the integration variables, the 
thus-obtained singlet and octet correction factors will not necessarily differ
by the factor $-8$.

At this point we stress that any observable that is inclusive in both 
top-quark invariant masses, such as the total cross-section, will not receive 
any non-factorizable corrections. This is a typical feature of these
interconnection effects~\cite{nf-cancellations}. As an example of a 
distribution that is subject to non-vanishing non-factorizable
corrections we focus on the invariant-mass distribution of the top quark,
which can be used for the mass determination. To this end we determine the 
non-factorizable correction $\delta_{\sss{nf}}(M)$ for the distribution
\be
       \frac{d\sigma}{dM} = 
       \frac{d\sigma_{\sss{Born}}}{dM}
       \Bigl[1+\delta_{\sss{nf}}(M)\Bigr],
\ee
where $M$ is the invariant mass of the $b$-quark and the $W^{+}$ boson.
The maximum of the Breit--Wigner distribution can be used to determine 
the top-quark mass. 
The linearized shift of this maximum as induced by the non-factorizable 
corrections is given by
\be
        \Delta M = \frac{1}{8} \, \Gamma_{t}^{2} \, 
        \frac{d\delta_{\sss{nf}}(M)}{dM}\Biggl|_{M=m_{t}}.
\ee

The correction $\delta_{\sss{nf}}(M)$ is calculated for the four different 
mechanisms of $t\bar{t}$ production, i.e.~initiated by $e^{+}e^{-}$, 
$\gamma\gamma$, $q\bar{q}$ and $gg$. For the centre-of-mass energies of these 
(partonic) reactions we take $\sqrt{s}=355\GeV$ and $500\GeV$. 
These values exemplify the non-factorizable corrections in the vicinity of the 
threshold and far above it. As mentioned before, the adopted approximation
in our calculation  (LPA) forces us to stay sufficiently far above the
$t\bar{t}$ threshold (read: a few times $\Gamma_{t}$). The numerical values 
for the input parameters are
\be         
         m_{t}=173.8 \GeV, \ \ \ M_{W} = 80.26 \GeV, 
         \ \ \ M_{Z} = 91.187\GeV,
\ee
and
\be
         \Gamma_{t} = 1.3901 \GeV,
\ee
the latter being the $\OO(\alpha_{\sss{S}})$ corrected top-quark width. The 
correction $\delta_{\sss{nf}}$ is proportional to $\alpha_{\sss{S}}$, for 
which we have to choose the relevant scale. For $\sqrt{s}=355\GeV$ the main 
contribution originates from the non-factorizable Coulomb effect present in 
$\delta_{\sss{nf}}$. Its typical momentum is determined by the top-quark width 
$\Gamma_{t}$ and velocity $\beta$: $\Gamma_{t}/\beta\sim 6.8\GeV$. 
At $500\GeV$ softer gluons contribute and therefore the typical gluon momentum 
is $\Gamma_{t}\sim 1.4\GeV$. Therefore we choose
\be
          \alpha_{\sss{S}}(1.4\GeV)\approx 0.3536
          \ \ \ 
          \mbox{for}
          \ \ \ 
          \sqrt{s}=500\GeV,
\ee
\be
          \alpha_{\sss{S}}(6.8\GeV) \approx 0.1955
          \ \ \ 
          \mbox{for}
          \ \ \ 
          \sqrt{s}=355\GeV,
\ee
corresponding to $\alpha_{\sss{S}}(M_{Z})=0.1180$ at the $Z$ peak.
It should be noted that choosing another scale in $\alpha_{\sss{S}}$
will only affect the normalization
of the correction. 
\begin{figure}[htb]
  \unitlength 0.7cm
  \begin{center}
  \begin{picture}(13.4,6.8)
  \put(1.6,5.9){\makebox[0pt][c]{\bf\boldmath$\delta_{\sss{nf}}$}} 
  \put(1.6,4.8){\makebox[0pt][c]{\bf [\%]}}
  \put(12.3,-0.3){\makebox[0pt][c]{\boldmath $M$ \bf [GeV]}}
  \put(1,-5.5){\includegraphics{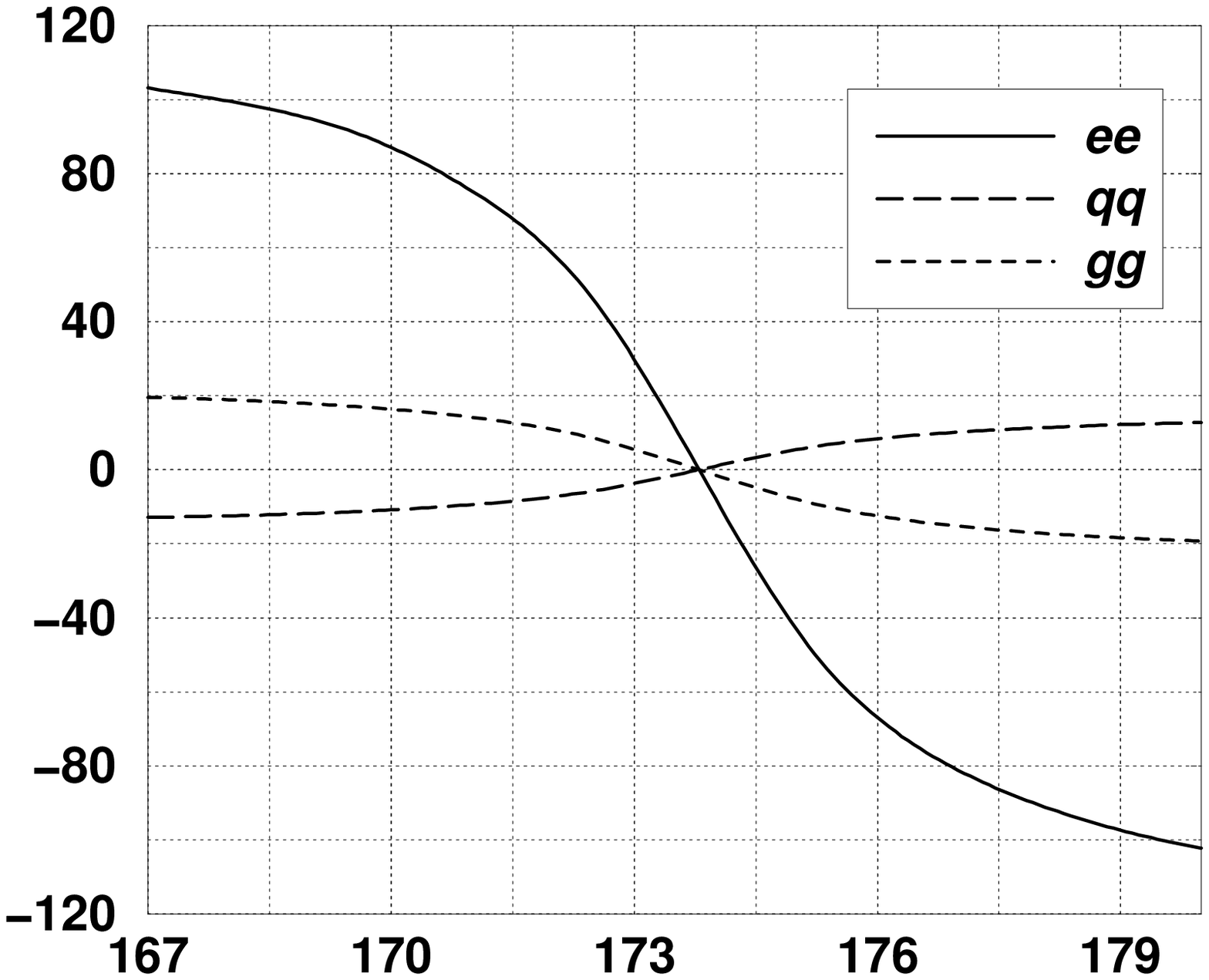}}
  \end{picture}
  \end{center}
  \caption[]{The relative non-factorizable correction $\delta_{nf}(M)$ to the 
             single invariant-mass distribution $d\sigma/dM$.
             Centre-of-mass energy: $\sqrt{s}=355~\GeV$.}
\label{fig:plot1} 
\end{figure}%
\begin{figure}[htb]
  \unitlength 0.7cm
  \begin{center}
  \begin{picture}(13.4,6.8)
  \put(1.6,5.9){\makebox[0pt][c]{\bf\boldmath$\delta_{\sss{nf}}$}} 
  \put(1.6,4.8){\makebox[0pt][c]{\bf [\%]}}
  \put(12.3,-0.3){\makebox[0pt][c]{\boldmath $M$ \bf [GeV]}}
  \put(1,-5.5){\includegraphics{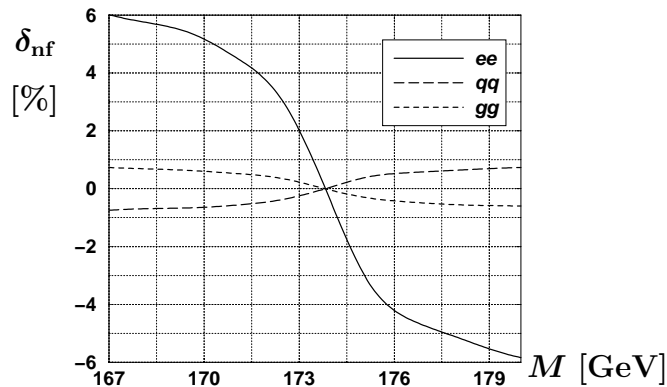}}
  \end{picture}
  \end{center}
  \caption[]{The relative non-factorizable correction $\delta_{nf}(M)$ to the 
             single invariant-mass distribution $d\sigma/dM$.
             Centre-of-mass energy: $\sqrt{s}=500~\GeV$.}
\label{fig:plot2}
\end{figure}%

In Fig.~\ref{fig:plot1} the non-factorizable correction
$\delta_{\sss{nf}}$ is plotted as a function of the invariant mass $M$ at the 
centre-of-mass energy of $355\GeV$. The $\delta_{\sss{nf}}$ values
for the pure singlet $e^{+}e^{-}$ initial state and the pure octet $q\bar{q}$ 
initial state differ approximately by the afore-mentioned factor of $-8$.
For the $gg$ initial state the Born octet part is larger than the singlet one,
resulting in a non-factorizable correction that lies between the $e^{+}e^{-}$
and the $q\bar{q}$ case.
The correction for the $\gamma\gamma$ initial 
state is virtually indistinguishable from the $e^{+}e^{-}$ one and is 
therefore not displayed. Evidently the distortion effects from the singlet 
corrections are very large, which is due to a large non-factorizable Coulomb 
correction inside $\delta_{\sss{nf}}$. The maximum of the Breit--Wigner 
distribution is hardly affected by this large correction.
One finds for the various initial states $e^{+}e^{-}(\gamma\gamma)$,
$gg$ and $q\bar{q}$ $\Delta M\approx -85$, $-15$  and $+10\MeV$
respectively.
The situation at $500\GeV$ is depicted in Fig.~\ref{fig:plot2}.
The overall correction is small, which is typical for non-factorizable
corrections further away from threshold. The shift in the maximum of the 
Breit--Wigner distribution is of the order of $5\MeV$ for 
the $e^{+}e^{-}$ and $\gamma\gamma$ initial states, and even smaller for the 
$q\bar{q}$ and $gg$ initial states.

In order to obtain hadronic distributions from the partonic ones, the results
for the $q\bar{q}$ and $gg$ initial states should of course be properly folded
with the parton densities of the colliding hadrons ($p\bar{p}$ at the Tevatron,
$pp$ at the LHC). The bulk of the partonic contributions originates from the 
energy region not far above the $t\bar{t}$ threshold ($s\lsim 8m_{t}^{2}$, 
i.e.~$\sqrt{s}\lsim500\GeV$), which is exemplified by the partonic energies
$355$ and $500\GeV$ used in our analysis.

\section{Conclusions}

In this paper we have summarized the gauge-invariant description for 
calculating the $\OO(\alpha_{\sss{S}})$ non-factorizable QCD corrections to 
pair production of top quarks. The formalism is presented in a general way, 
making it applicable to all relevant initial states. The resulting final 
formula for the non-factorizable corrections involves the same quantity 
$\delta_{\sss{nf}}$ for all reactions. This quantity can be numerically 
calculated using expressions available in the literature.

Although the formalism can be used for numerical studies of many distributions,
the focus of our numerical evaluation has been on the invariant-mass 
distribution of the top quark, which can be used for extracting the top-quark 
mass. In spite of the possible sizeable deformations of this line-shape 
distribution, its maximum is shifted by less then $100\MeV$. 
Therefore, if the top-quark mass is extracted experimentally from the peak 
position of the line-shape, the non-factorizable corrections can be safely 
neglected. If the precise shape of the Breit--Wigner distribution is
used in the experimental analysis, the non-factorizable corrections
should be taken into account properly. In particular if the singlet
colour state dominates. In addition higher-order non-factorizable
corrections might be needed.
\\
\\
{\bf Acknowledgement}\\
Discussions with V.A.~Khoze are gratefully acknowledged.

\end{document}